\documentclass[10pt,journal,compsoc]{IEEEtran}
\ifCLASSOPTIONcompsoc
  \usepackage[nocompress]{cite}
\else
  \usepackage{cite}
\fi
\ifCLASSINFOpdf
\else
\fi

\usepackage{rotating}
\usepackage{lscape}
\usepackage{caption}
\usepackage{subcaption}
\usepackage{multirow}
\usepackage{booktabs}
\usepackage{rotating}
\usepackage{lscape}
\usepackage{caption}
\usepackage{threeparttable}

\usepackage[table,x11names]{xcolor}

\usepackage[most]{tcolorbox}
\usepackage{lipsum}
\usepackage[bordercolor=white,backgroundcolor=gray!30,linecolor=black,colorinlistoftodos]{todonotes}

\usepackage{xcolor, mdframed}
\usepackage[normalem]{ulem} 
\newcommand\hll{\bgroup\markoverwith
  {\textcolor{yellow}{\rule[-.5ex]{2pt}{2.5ex}}}\ULon}

\begin{document}
%
\title{A Survey of System Architectures and Techniques for FPGA Virtualization}
%
%
%
%

\author{Masudul Hassan Quraishi,
        Erfan Bank Tavakoli,
        Fengbo Ren
\IEEEcompsocitemizethanks{\IEEEcompsocthanksitem M. Quraishi, E. Bank Tavakoli, F. Ren are with the School of Computing, Informatics, and Decision Systems Engineering, Arizona State University, Tempe,
AZ, 85281.\protect\\
E-mail: mquraish@asu.edu; ebanktav@asu.edu; renfengbo@asu.edu.
}
}

\IEEEtitleabstractindextext{%
\begin{abstract}
FPGA accelerators are gaining increasing attention in both cloud and edge computing because of their hardware flexibility, high computational throughput, and low power consumption. However, the design flow of FPGAs often requires specific knowledge of the underlying hardware, which hinders the wide adoption of FPGAs by application developers. Therefore, the virtualization of FPGAs becomes extremely important to create a useful abstraction of the hardware suitable for application developers. Such abstraction also enables the sharing of FPGA resources among multiple users and accelerator applications, which is important because, traditionally, FPGAs have been mostly used in single-user, single-embedded-application scenarios. There are many works in the field of FPGA virtualization covering different aspects and targeting different application areas. In this survey, we review the system architectures used in the literature for FPGA virtualization. In addition, we identify the primary objectives of FPGA virtualization, based on which we summarize the techniques for realizing FPGA virtualization. This survey helps researchers to efficiently learn about FPGA virtualization research by providing a comprehensive review of the existing literature. 
\end{abstract}

}

\maketitle

\IEEEdisplaynontitleabstractindextext

%
\IEEEpeerreviewmaketitle

\IEEEraisesectionheading{\section{Introduction}\label{sec:introduction}}

%
%
%
%
\IEEEPARstart{F}{ield} programmable gate arrays (FPGAs) are gaining increasing attention in both cloud and edge computing because of their hardware flexibility, superior computational throughput, and low energy consumption \cite{biookaghazadeh2018fpgas}. Recently, commercial cloud services, including Amazon \cite{amazonf1} and Microsoft \cite{catapult2014}, have been employing FPGAs. In contrast with CPUs and GPUs that are widely deployed in the cloud, FPGAs have several unique features rendering them synergistic accelerators for both cloud and edge computing. First, unlike CPUs and GPUs that are optimized for the batch processing of memory data, FPGAs are inherently efficient for processing streaming data from inputs/outputs (I/Os) at the network edge. With abundant register and configurable I/O resources, a streaming architecture can be implemented on an FPGA to process data streams directly from I/Os in a pipelined fashion. The pipeline registers allow efficient data movement among processing elements (PEs) without involving memory access, resulting in significantly improved throughput and reduced latency \cite{tessier2001reconfigurable, woods2008fpga}. Second, unlike CPUs and GPUs that have a fixed architecture, FPGAs can adapt their architecture to best fit any algorithm characteristics due to their hardware flexibility. Specifically, the hardware resources on an FPGA can be dynamically reconfigured to compose both spatial and temporal (pipeline) parallelism at a fine granularity and on a massive scale \cite{biookaghazadeh2018fpgas,markovic2012dsp}. As a result, FPGAs can provide consistently high computational throughput for accelerating both high-concurrency and high-dependency algorithms, serving a much broader range of cloud and edge applications. Third, FPGA devices consume an order of magnitude lower power than CPUs and GPUs and are up to two orders of magnitude more energy-efficient, especially for processing streaming data or executing high-dependency tasks \cite{vestias2014trends,bank2019polar,dorrance2014scalable}. Such merits lead to improved thermal stability as well as reduced cooling and energy costs, which is critically needed for both cloud and edge computing.

Even though FPGAs offer great benefits over CPUs and GPUs, these benefits come with design and usability trade-offs. Conventionally, FPGA application development requires the use of a hardware description language (HDL) and knowledge about the low-level details of FPGA hardware. This is a deal-breaker for most software application developers that are not familiar with HDLs nor hardware specifics at all. Even though high-level synthesis (HLS) has enabled the development of FPGA kernels in C-like high-level languages (e.g., C++/OpenCL)\cite{munshi2009opencl}, one still needs to have basic knowledge about FPGA hardware specifics in order to develop performance-optimized FPGA kernels. As a result, the FPGA application development based on HLS remains esoteric. Moreover, the existing flows of FPGA kernel design are highly hardware-specific (each FPGA kernel binary is specific to one FPGA model only), and the vendor-provided tools for deploying and managing FPGA kernels lack the support for sharing FPGA resources across multiple users and applications. These limitations make FPGAs insufficient for supporting multi-tenancy cloud and edge computing. One solution to these limitations is decoupling the application (\textit{i.e.,} hardware-agnostic) and the kernel design (\textit{i.e.,} hardware-specific) development regions \cite{riera2020halo}.

FPGA virtualization that aims to address the challenges mentioned above is the key to enabling the wide adoption of FPGAs by software application developers in multi-tenancy cloud and edge computing. Specifically, the primary objectives of FPGA virtualization are to: 1) create abstractions of the physical hardware to hide the hardware specifics from application developers and the operating system (OS) as well as provide applications with simple and familiar interfaces for accessing the virtual resources; 2) enable the efficient space- and time-sharing of FPGA resources across multiple users and applications to achieve high resource utilization; 3) facilitate the transparent provisioning and management of FPGA resources; and 4) provide strict performance and data isolation among different users and applications to ensure data security and system resilience.

An early survey paper on virtualization of reconfigurable hardware \cite{Plessl2004VirtualizationOH} covers a narrow perspective of FPGA virtualization by discussing three virtualization approaches, i.e., temporal partitioning, virtualized execution, and virtual machine. Recent surveys on FPGA virtualization \cite{Vaishnav2018ASO, ijaz2020revisiting, skhiri2019fpga} mostly discuss FPGA virtualization in the context of cloud and edge computing. Vaishnav et al. \cite{Vaishnav2018ASO} proposed three categories of FPGA virtualization techniques and approaches based on the level of abstraction and the scale of computing resource deployment: resource level, node level, and multi-node level. The resource-level abstraction refers to the virtualization of reconfigurable hardware and I/O resources on an FPGA, while the node-level and multi-node-level abstractions refer to the virtualization of a computing system with one and multiple FPGAs, respectively. Such categorization can create ambiguity since many approaches and techniques used in one abstraction level of the system can often be applied to other levels as well. For example, the paper considers scheduling as a node-level virtualization technique. The scheduling of FPGA tasks is, in fact, a common concept that can also be applied to the multi-node and resource levels. Furthermore, the survey lists several FPGA virtualization objectives, but a discussion on how these objectives are linked to the virtualization techniques in the existing literature is missing. 

The survey paper in \cite{ijaz2020revisiting} revisits some of the existing survey papers to suggest selection criteria of appropriate communication architectures and virtualization techniques for deploying FPGAs to data centers. It has selected three different areas to review: previously used nomenclature, Network-on-Chip (NoC) evaluation on FPGA, and FPGA virtualization. Its discussions in the FPGA virtualization section are similar to \cite{Vaishnav2018ASO}, and therefore have the same limitations mentioned above. 

Skhiri et al. \cite{skhiri2019fpga} identified drawbacks of using FPGA with a local machine and presented a survey of papers that uses FPGAs in the cloud to solve the identified drawbacks. They discussed cloud FPGA services in three different groups: software tools, platforms, and resources. As part of their discussion on FPGA-resources-as-a-service, the paper revisits the classification of virtualization approaches in \cite{Vaishnav2018ASO}, and hence, has the same limitations. 

Overall, the existing survey papers on FPGA virtualization review a limited set of existing work; for example, literature discussing isolation and security perspectives of FPGA virtualization are not reviewed. There is an existing survey on security and trust of general FPGA-based systems \cite{zhang2014survey}, but the perspective of virtualization is not covered in the work. One of the surveys has overlapped categorization, which creates ambiguity in the scope of which each virtualization technique can apply. The existing surveys also fall short in linking the virtualization techniques to the core objectives of FPGA virtualization. In addition, the existing surveys fail to discuss the system architecture perspective of FPGA virtualization techniques. As FPGAs are reconfigurable hardware, FPGA computing systems can be built with a variety of system architecture choices, and the virtualization techniques can be highly dependant on the architecture choices. Thus, reviewing the system architectures of FPGA computing systems is critical to understanding the corresponding FPGA virtualization techniques.

In this survey paper, we present a comprehensive review of both the system architectures, covering the hardware, software, and overlay stacks, and the techniques for FPGA virtualization, as well as their associations. Such organization of the survey sets a clear boundary among different system stacks, which helps readers better understand how different virtualization techniques apply to different system architectures. Furthermore, our work elaborates the four key objectives of FPGA virtualization by discussing how each objective is addressed by different virtualization techniques in the literature. The system architectures used for FPGA virtualization are summarized in Table \ref{tab:table-1}, and the key techniques for realizing FPGA virtualization are summarized in Table \ref{tab:table-3}, categorized by the four primary objectives of FPGA virtualization. In addition, this survey paper also reviews the existing papers on the isolation and security issues of multi-tenant FPGAs that are overlooked by the previous surveys. 

This survey paper is organized as follows. Section \ref{sec:background} provides background on the general concepts of virtualization and the challenges of FPGA virtualization, clarifies the important definitions used throughout this paper, and reviews the available programming models adopted in FPGA virtualization. Section \ref{sec:system architecture} presents the system architecture design for FPGA virtualization in detail. In Section \ref{sec:virtualization objectives}, we discuss the four objectives of FPGA virtualization and how they are implemented in different system stacks. Section \ref{sec:conclusion} summarizes the overall survey and draws the conclusion. 

\section{Background}
\label{sec:background}

\subsection{Virtualization: General Concepts}
\label{subsec2.1:Virtualization General }
Virtualization, in general, is creating a virtual abstraction of computing resources to hide the low-level hardware details from users. Virtualization is often software-based, \textit{i.e.,} the virtual abstraction is implemented on the software level to hide the complexity of the underlying hardware. In virtualization, the resources are transparently presented to users so that each user has an illusion of having unlimited and exclusive access. The most common example of virtualization is running different OSs on a single processor using virtual machines, where a virtual machine creates the illusion of a standalone machine with an OS to the users.  In this way, the users get the flexibility to easily switch to a different system environment or even a different OS without changing the computing hardware. 

For GPU virtualization, in gVirtuS \cite{di2012virtualizing}, a split driver approach is utilized with a frontend component (guest-side software component) deployed on virtual machine images and a backend component that manages device requests and accesses and multiplexes devices. In \cite{duato2011enabling}, rCuda is proposed to access GPUs on HPC clusters within virtual machines remotely. The rCuda framework has two components:  1) the client middleware consisting of a collection of wrappers in charge of forwarding the API calls from the applications requesting acceleration services to the server middleware and retrieving the results back, and 2) the server middleware that receives, interprets, and executes the API calls from the clients and performing GPU multiplexing.

Similar to the general concept of virtualization, FPGA virtualization is creating a virtual abstraction of FPGA resources to hide the intricate low-level hardware details from users or application developers. The definition of FPGA virtualization has changed significantly over time. The early work on FPGA virtualization in the 90s \cite{VirtualFPGA1998} introduced OS principles, such as partitioning, segmentation, and overlaying for FPGAs and termed these techniques as the virtualization techniques of FPGAs. Later on, the work in the 00s \cite{Placing-Routing-2001} defined FPGA virtualization as an abstract hardware layer on top of physical FPGA fabrics. This layer is now commonly known as the overlay architecture \cite{So2016}. In \cite{Plessl2004VirtualizationOH}, three approaches for FPGA virtualization are presented: temporal partitioning, virtualized execution, and mapping to an abstract virtual machine. As FPGA technology has evolved dramatically over the past few decades, the goal of FPGA virtualization has also changed over time. Today, from the perspective of cloud and edge computing, the primary objectives of FPGA virtualization are to create a useful abstraction of the underlying hardware to facilitate application developers and users and enable the efficient sharing and management of FPGA resources across multiple users and applications with strict performance and data isolation.

In a recent survey on FPGA Virtualization \cite{Vaishnav2018ASO}, the objectives of FPGA virtualization are listed as: \textit{multi-tenancy}, \textit{resource management}, \textit{flexibility}, \textit{isolation}, \textit{scalability}, \textit{performance}, \textit{security}, \textit{resilience}, and \textit{programmer's productivity}. However, we believe that scalability, performance, and resilience are the primary considerations of cloud and edge computing in general but not specific to FPGA virtualization, even though virtualization may facilitate improvements on some of those perspectives. Also, flexibility and programmer's productivity are the primary objectives of HLS techniques rather than FPGA virtualization. Therefore, we define the primary objectives of FPGA virtualization in the context of cloud and edge computing as the following:

\begin{itemize}
\item \textbf {Abstraction:} Abstraction hides the hardware specifics from application developers and the OS as well as provide applications with simple and familiar interfaces to access the virtual FPGA resources. 
\item \textbf {Multi-tenancy:} To enable the efficient sharing of FPGA resources among multiple users and applications. 
\item \textbf{Resource Management:} To facilitate the transparent provisioning and management of FPGA resources for workload balancing and fault tolerance.
\item \textbf{Isolation:} To provide strict performance and data isolation among different users and applications to ensure data security and system resilience.
\end{itemize}

Our discussion on the FPGA virtualization techniques in the paper will focus on these four objectives. While the previous survey paper only listed the objectives of FPGA virtualization, our discussion in Section \ref{sec:virtualization objectives} elaborates on how each of the four key objectives is addressed by different virtualization techniques in the existing literature as well as their association with different stacks of the system architecture.

\subsection{Challenges of FPGA Virtualization}
\label{subsec2.2:Challenges}
It should be noted that virtualization is not hardware-agnostic. One must know about the hardware specifics of a computing system in order to properly virtualize it. In the case of FPGAs, the underlying hardware is reconfigurable, and thus the hardware architecture can vary according to the application implemented. Such hardware flexibility makes the virtualization of FPGA resources much more challenging than the virtualization of CPUs and GPUs with a fixed hardware architecture. 

Traditionally, FPGAs are used primarily for embedded applications, where a dedicated application running on an FPGA is only accessible to a dedicated user. Although modern commercial FPGAs have the potential to support multiple applications via partial reconfiguration (PR), the existing FPGA architectures and design flows are still not optimized for sharing hardware resources among multiple users and applications, making the multi-tenancy computing objective of FPGA virtualization a challenging task to accomplish.  

In addition, FPGAs have limited performance portability. Since FPGA architectures and design flows are both vendor- and hardware-specific, it is impossible to apply the same FPGA kernel (bitfile) to a different FPGA device offered by the same or a different vendor.  Although HDL and HLS codes are portable across devices and vendors (other than vendor-provided IP cores and language directives), the re-compilation needed for mapping the same application codes onto different FPGA devices can be extremely time-consuming, creating difficulties for the run-time deployment and provisioning of FPGA applications in a cloud or edge computing environment.  Due to the tight coupling between FPGA hardware and vendor-specific design flows, the development of a generalized framework for FPGA virtualization becomes a challenge.  This makes the transparent FPGA resource management objective of FPGA virtualization a challenging task to accomplish.

\subsection{Definitions}
\label{subsec2.3:Definition}
We found that the use of terminologies for FPGA virtualization in the existing literature is not always consistent. Sometimes, different terminologies are used to refer to the same or similar system architecture components. To avoid confusion and ambiguity, we uniformly define three important terminologies for FPGA virtualization as follows.

\begin{itemize}
\item \textit{Shell} refers to the static region of FPGA hardware, which is pre-designed and pre-implemented prior to application deployment and fixed at run time. Common glue logic is packaged into a reusable shell framework to enhance hardware re-usability. No shell framework is able to fulfill the requirements of all applications. Introducing more functionalities into a shell framework to cover more application requirements also results in increased resource usage, higher power consumption, and possibly lower clock frequency of the static region of FPGA hardware. In some of the literature, a shell is also referred to as a hardware OS \cite{Feniks2017,ResourceElastic2018}, a static region/part \cite{RC2F2017,Ker-ONE2016}, and hull \cite{khawaja2018sharing}.
\item \textit{Role} refers to the dynamic region of FPGA hardware, which can be independently reconfigured to map different applications at run time. A role is commonly comprised of several partially reconfigurable regions (PRRs). Each of the PRRs can be reconfigured on-the-fly without interrupting the operation of other PRRs using the technique known as dynamic partial reconfiguration (DPR)\cite{DPR2009}. In some of the literature, a role is referred to as an application region \cite{FPGAs-Cloud2017}, a reconfiguration region \cite{VFR2014}, or global zone \cite{khawaja2018sharing}.
\item \textit{vFPGA} or virtual FPGA is a virtual abstraction of the physical FPGA. One or multiple PRRs can be mapped to a single vFPGA. 
\item \textit{Accelerator} refers to a PRR or vFPGA programmed with an application. 
\item \textit{Hypervisor} creates an abstraction layer between software applications and the underlying FPGA hardware and manages the vFPGA resources. In some of the literature, a hypervisor is referred to as an OS, a resource management system/framework \cite{RC3E2016}\cite{VirtualRuntime2016}, a virtual machine monitor (VMM) \cite{pvFPGA2013}, a run-time manager \cite{VirtualRuntime2016}, a hardware task manager \cite{HWSWVDI2013}, a context management \cite{NFVManager2017}, or a virtualization manager \cite{HPRCVCM2012}.
\end{itemize}

\subsection{Programming Model}
\label{subsec2.4:Programming Model}
There are three programming models for application development supported by the existing FPGA virtualization frameworks: domain specific language (DSL) programming, HDL programming at the register transfer level (RTL), and HLS programming. 

In DSL programming, the applications are written in a high-level language (\textit{e.g.}, C). Selected portions of codes (\textit{e.g.}, loops) are separated and then implemented as FPGA kernels through the conventional RTL or HLS design flows. For example, in \cite{QuickDough2015}, loops are identified, and a data flow graph is created. The data flow graph is then converted into an FPGA kernel through the conventional RTL flow. 

In HDL programming, FPGA kernels are designed using an HDL. Common HDLs include Verilog, SystemVerilog, and VHDL. HDLs are the most commonly used programming languages in FPGA Virtualization, as it is the traditional programming language for embedded FPGA design. Writing HDL codes for achieving optimized resource utilization, power consumption, and performance requires significant design effort as well as expert knowledge about hardware design, which makes it difficult for software application developers to master. From Table \ref{tab:table-1}, we observe that most of the prior work use HDL programming as the programming model, which unfortunately is a deal-breaker for the adoption of FPGAs by software application developers in cloud and edge computing. 

In HLS programming, a high-level programming language (\textit{e.g.}, OpenCL\cite{munshi2009opencl}, Java\cite{pell2013maximum, pell2012maximum}) is used to develop FPGA kernels \cite{li2018gpu}. Then, the high-level language codes are translated into HDL codes. Different from HDL programming, HLS programming requires less knowledge about the underlying hardware and provides a significant improvement in design productivity. Some HLS-based design tools provide not only compilation and simulation utilities but also run-time utility for FPGA management. Take Intel FPGA SDK for OpenCL\cite{munshi2009opencl} as an example. It provides an offline compiler for generating FPGA kernel hardware with interface logic, a software-based emulator for symbolic debugging and functional verification, and a run-time environment including firmware, software, and the device driver to control and monitor the kernel deployment and execution, and transfer data between the system memories of the host and the FPGA computing device.

\section{System Architecture for FPGA Virtualization}
\label{sec:system architecture}

\subsection{Hardware Stack}
\label{subsec3.1:Hardware Stack}
The hardware stack of FPGA virtualization varies in the design of three components: 1) the host interface that defines how an FPGA computing device is connected to a host CPU; 2) the shell that defines the storage, communication, and management capability of an FPGA computing device; and 3) the role that defines the run-time application mapping capability of an FPGA computing device. 

\begin{figure}[]
\centering
\begin{minipage}{.1\textwidth}
  \includegraphics[width=1\linewidth]{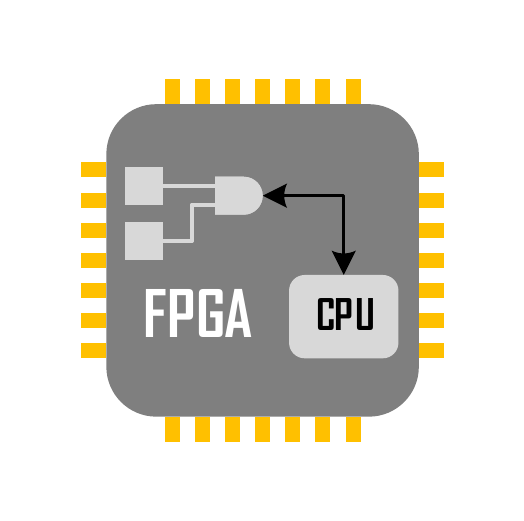}  
  \subcaption{On-Chip}
  \label{fig:On-Chip}
\end{minipage}
\begin{minipage}{.2\textwidth}
  \includegraphics[width=1\linewidth]{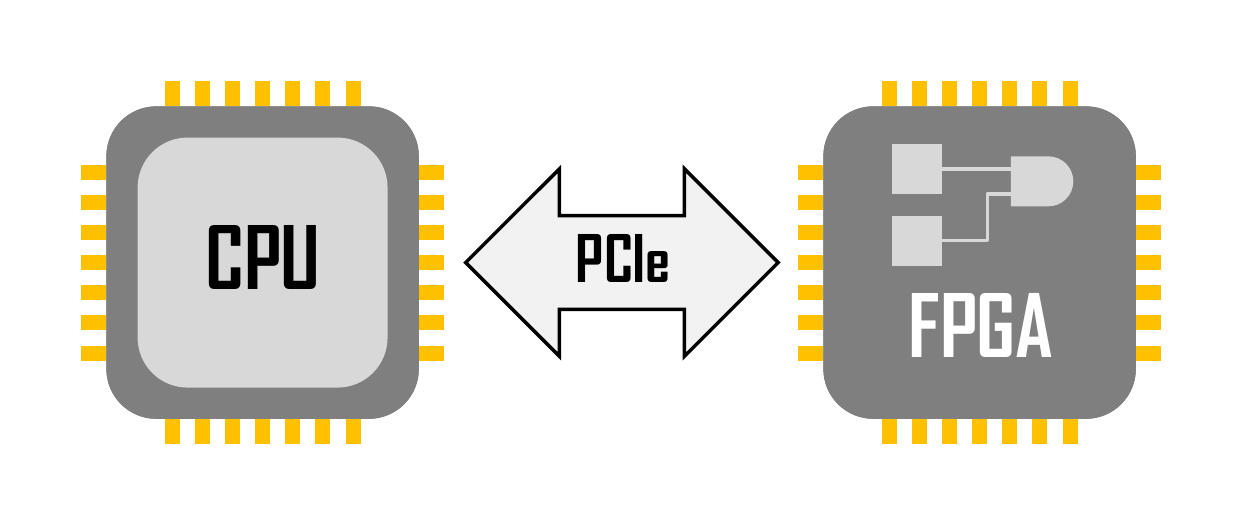}  
  \subcaption{Local Host}
  \label{fig:Local Host}
\end{minipage}
\begin{minipage}{.2\textwidth}
  \includegraphics[width=1\linewidth]{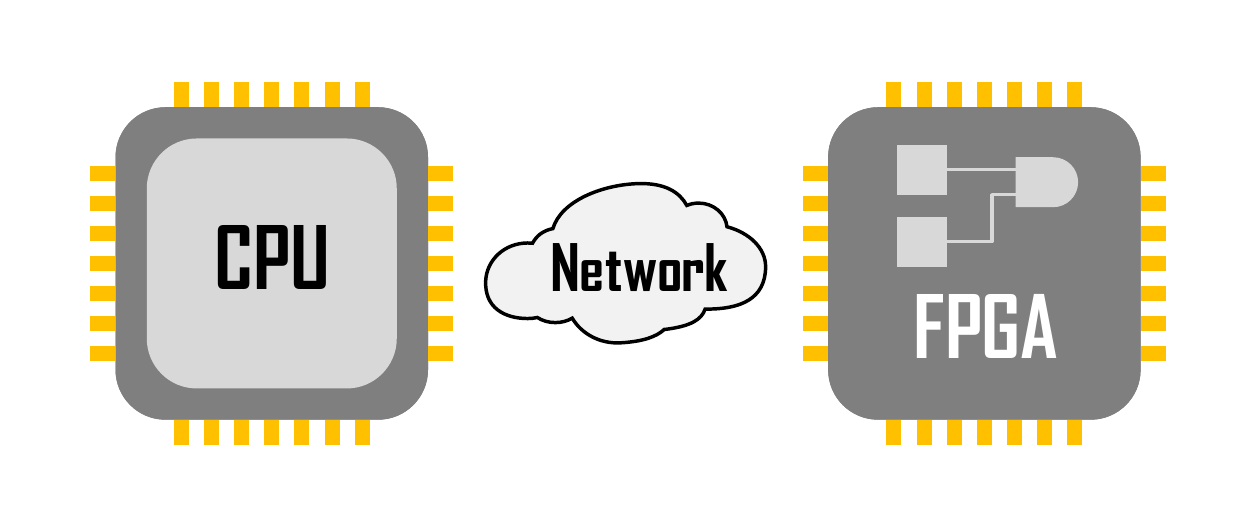}  
  \subcaption{Remote Host}
  \label{fig:Remote Host}
\end{minipage}
\caption{Host interfaces for FPGAs}
\label{fig:Connection}
\end{figure}
\subsubsection{Host Interface}
There are three different types of host interfaces available for the existing FPGA devices (Fig. \ref{fig:Connection}) described as follows.

\begin{itemize}
    \item \textbf{On-Chip Host Interface} connects the FPGA computing hardware to a dedicated soft/hard CPU implemented on the same chip (\textit{e.g.}, a Xilinx MicroBlaze soft processor or an ARM-based hard processor on Xilinx Zynq FPGAs) (see Fig. \ref{fig:On-Chip}) \cite{OS4RS2009}. The on-chip host interface has the lowest communication latency, thus has the highest performance among all the interface types. However, a hard CPU implementation occupies the FPGA chip area, and a soft CPU implementation consumes the reconfigurable resources available on an FPGA device, which reduces the capacity of application mapping.
\end{itemize}

\begin{landscape}
\centering
\begin{table}
\scriptsize
\caption{\label{tab:table-1} Summary of the System Architectures, Programming Models, Applications, and FPGA Platforms Adopted by the Existing FPGA Virtualization Work}
\begin{threeparttable}
\renewcommand{\arraystretch}{1.2}
\begin{tabular}{|c|c|p{4cm}|c|p{3cm}|p{2cm}|c|p{4cm}|p{2.3cm}|}
\hline
\multirow{3}{*}{\textbf{Work}} & \multicolumn{5}{c|}{\textbf{System Architecture}}                                                                                                                             & \multirow{3}{*}{\textbf{PM}} & \multirow{3}{*}{\textbf{Application}}                            & \multirow{3}{*}{\textbf{Platform}}                      \\ \cline{2-6}
                               & \multicolumn{3}{c|}{\textbf{Hardware}}                                                       & \multirow{2}{*}{\textbf{Software}} & \multirow{2}{*}{\textbf{Overlay}}         &                              &                                                                  &                                                         \\ \cline{2-4}
                               & \textbf{Host}  & \textbf{Shell}                                        & \textbf{Role} &                                    &                                           &                              &                                                                  &                                                         \\ \hline \hline
Virtual \cite{VirtualFPGA1998, GeneralVirtualization1998} & N/A & N/A & N/A & N/A & N/A & N/A & N/A & N/A \\ \hline
Firm-core JIT \cite{Firm-coreJIT2005} & N/A & N/A & N/A & N/A & CGRA & HDL & MCNC Benchmark & Xilinx Spartan-IIE \\ \hline
Cray HPRC \cite{CrayHPRC2008} & LH & None & PRR & N/M & N/A & HDL & Image Feature Extraction & Cray XD1 \\ \hline
OS4RS \cite{OS4RS2009} & OC & Communication Component & PRR & OS4RS & N/A & HDL & Multimedia & Xilinx Virtex-II \\ \hline
NoC \cite{HomogeneousNoC2010} & LH & N/A & N/A & N/M & HRCs with NoC & N/A & 3-DES Crypto Algorithm & Xilinx Virtex-5 \\ \hline
VirtualRC \cite{VirtualRC2012} & LH & N/A & N/A & Software-Middleware API & N/A & HLS, HDL & Bioinformatics & GiDEL, Nallatech , Pico Computing \\ \hline
HPRC \cite{HPRCVCM2012} & LH & N/A & N/A & N/M & N/A & HDL & IDEA Cipher/Euler CFD Solver & Altera Stratix III \\ \hline
pvFPGA \cite{pvFPGA2013} & LH & PCIe Controller, DMA & N/A & Xen VMM & N/A & HDL & FFT & Xilinx Virtex-6 \\ \hline
VDI \cite{HWSWVDI2013} & OC & Memory/Configuration/Network Controller & PRR & N/M & N/A & N/A & Multimedia & Xilinx Virtex-II \\ \hline
ACCL \cite{AcceleratorPool2014} & LH & PCIe Controller, DMA & PRR & OpenStack & N/A & N/A & Cryptography & Xilinx Kintex-7 \\ \hline
VFR \cite{VFR2014} & LH & NetFPGA BSP & PRR & OpenStack, SAVI Testbed & N/A & HLS & Load Balancer & Xilinx Virtex-5 \\ \hline
Catapult \cite{catapult2014} & LH, RH & Memory Controller, Host Comm., Network Access & N/A & N/M & N/A & HDL & Ranking Portion of the Bing Web Search Engine & Intel Stratix V \\ \hline
FF HLS\cite{FastFlexibleHLS2014} & LH & N/A & N/A & OpenCL & Intermediate Fabrics & HLS & Computer Vision and Image Processing & Xilinx Virtex-6 \\ \hline
HSDC \cite{HSDC2015} & OC, RH & Management Layer, Network Service Layer & PRR & OpenStack & N/A & HDL & N/M & Xilinx Zynq-7100 \\ \hline
VFA Cloud \cite{VFACloud2015} & LH & PCIe Controller, DMA & PRR & DyRACT Test Platform & N/A & N/A & Map Reduce & Xilinx Virtex-7 \\ \hline
RC3E \cite{RC3E2016} & LH, RH & Communication/Virtualization Infrastructure & PRR & N/A & N/A & HDL & Black-Scholes Monte Carlo Simulation & Xilinx Virtex-7 \\ \hline
RRaaS \cite{RRaaS2016} & LH & PCIe Controller & PRR & N/A & Soft Processor & HDL & Image Processing & N/M \\ \hline
Virt. Runtime \cite{VirtualRuntime2016} & LH & PCIe Controller, Run-time Manager & PRR & Multi-Threaded OS & N/A & DSL, HLS, HDL & Graph Algorithms: Page Rank, Triangle Count and Outlier Detector & Xilinx Virtex-7 \\ \hline
Ker-ONE\cite{Ker-ONE2016} & OC & N/A & PRR & RTOS & N/A & N/A & DSP & Xilinx ZedBoard \\ \hline
RC2F \cite{RC2F2017} & LH, RH & Hypervisor, I/O Components & DPRR & N/A & N/A & HDL & N/M & Xilinx Virtex-7 \\ \hline
VER Cloud \cite{VERCloud2017} & OC, LH & PCIe Controller, Run-time Manager & PRR & Multi-Threaded OS & N/A & DSL, HLS, HDL & Graph Algorithms: Page Rank, Triangle Count and Outlier Detector & Xilinx Virtex-7 \\ \hline
Feniks \cite{Feniks2017} & RH & Host Comm., Netwtork/Storage Stack, Memory Controller & PRR & N/M & N/A & HDL & Data Compression Engine, Network Firewal & Intel Stratix V \\ \hline
Cost of Virt. \cite{CostofVirtualization2017} & LH & DDR3/PCIe Controller, Ethernet Core & N/A & N/A & NoC Emulation & HDL & DSP & Intel Arria 10 \\ \hline
NFV \cite{NFVManager2017} & OC, LH & PCIe Controller, DMA & PRR & N/M & MCU on Soft Processor & HDL & Context Switch & Xilinx Virtex-7 \\ \hline
FPGAs-Cloud \cite{FPGAs-Cloud2017} & LH & Memory Controller & PRR & OpenCL & N/A & HLS & Network Functions & Xilinx Virtex-7 \\ \hline
RACOS \cite{RACOS2017} & LH & PCIe Controller, DMA, ICAP & DPRR & User API, Kernel driver & N/A & HDL, HLS & Edge and motion detection & Xilinx Virtex 6 \\ \hline
AmorphOS \cite{khawaja2018sharing} & LH & PCIe Controller, DMA, MMIO & DPRR & lib-AmorphOS & N/A & HDL & CNN, Memory streaming, Bitcoin & Altera Stratix V GS, Xilinx UltraScale+ \\ \hline
Elastic \cite{ResourceElastic2018} & OC & AXI & DPRR & OpenCL & N/A & HLS & Scheduling Algorithm & TE8080 Board \\ \hline
FPGAVirt \cite{VirtioHWSB2018, FPGAVirt2018} & LH & PCIe Controller & N/A & Virtio-Vsock & CGRA & HDL & NoC Emulation & Intel Stratix V \\ \hline
hCODE \cite{hCODE2018} & LH & PCIe Controller, Clock generator & PRR & N/M & N/A & HDL & Sorter, AES, KMeans & Xilinx Virtex-7/Kintex-7 \\ \hline
ViTAL \cite{zha2020virtualizing} & RH & Latency-insensitive Interface, Address Translation & DPRR & N/A & N/A & HLS & Machine Learning & Xilinx UltraScale+ \\ \hline
\end{tabular}
\renewcommand{\arraystretch}{1}
\begin{tablenotes}
\begin{tiny}
\item[\tnote{\textdagger}] 
N/A = Not Applicable; N/M = Not Mentioned; PM = Programming Model; LH = Local Host; RH = Remote Host; OC = On Chip
\end{tiny}
\end{tablenotes}
\end{threeparttable}
\end{table}
\end{landscape}

\begin{itemize}
    \item \textbf{Local Host Interface} is the most popular approach, which connects an FPGA computing device implemented as a daughter-card to a local CPU host through a high-speed serial communication (\textit{e.g.} bus, such as PCI Express (PCIe) (see Fig. \ref{fig:Local Host}) \cite{hCODE2018, khawaja2018sharing}.
    \item \textbf{Remote Host Interface} connects an FPGA computing device to a remote CPU host through the network, which allows the FPGA computing device to communicate with the CPU host node remotely (see Fig. \ref{fig:Remote Host}) \cite{Feniks2017, zeng2020enabling}. A remote host interface has the highest communication latency among all the interfaces. But, a remote host interface decouples an FPGA from a local host CPU, thus enables the integration of standalone FPGA computing devices and greatly improves the scalability of FPGA computing.
\end{itemize}

\subsubsection{Shell}
As mentioned in section \ref{subsec2.3:Definition}, a shell is the static region of FPGA hardware that contains the necessary glue logic and controllers for handling system memory access and the communications to a CPU host or other network devices. Thus, a shell should include but not limited to  a \textit{system memory controller}, a \textit{host interface controller}, and a \textit{network interface controller}.

\begin{itemize}
    \item \textbf{System Memory Controller} provides a user-friendly interface to the FPGA kernel for accessing the system memory of the FPGA computing device (\textit{e.g.}, a DDR SDRAM or high-bandwidth memory), which can be implemented either off-chip or on-chip (\textit{i.e.}, SoC design). In the existing literature, a system memory controller is also referred to as a DMA Engine \cite{Feniks2017}, DMA Controller \cite{VFACloud2015}, DRAM Controller \cite{VFR2014}, DRAM Adapter \cite{VFACloud2015}, DDR3 Controller \cite{CostofVirtualization2017}, or Memory Manager \cite{VERCloud2017}.
    \item \textbf{Host Interface Controller} enables the communication of an FPGA computing device with a CPU host through either an on-chip, local, or remote host interface. In the existing literature, a host interface controller is also referred to as a PCIe Module \cite{FPGAs-Cloud2017}, PCIe Controller \cite{NFVManager2017}, and DMA Controller \cite{pvFPGA2013}.
    \item \textbf{Network Interface Controller} provides a communication interface to a network (\textit{e.g.}, cloud network), which facilitates the communication of an FPGA computing device to other network devices without the intervention of the host CPU. In the existing literature, a network interface controller is also referred to as a Network Stack \cite{FPGAVirt2018}, Network Service Layer \cite{HSDC2015}, Ethernet Core and \cite{CostofVirtualization2017}.
\end{itemize}

\subsubsection{Role}
As mentioned in section \ref{sec:background}, a role is the dynamic region of FPGA hardware reserved for application mapping at run time. There are two approaches to reconfigure a role, namely, flat compilation and partial reconfiguration. In flat compilation, the whole role is reconfigured as a single region and is exclusive to single-accelerator applications. In partial reconfiguration, a role can contain multiple DPR regions (DPRRs), and a single DPRR can be reconfigured independently, or multiple DPRRs can be combined as one bigger DPRR and be reconfigured independently from the test PRRs at run time. The use of multiple DPRRs enables application deployment with flexible size of FPGA fabrics, thus can achieve higher resource utilization with improved overall system performance \cite{ResourceElastic2018}. Considering DPR, It is of great importance to reduce the reconfiguration time as much as possible using techniques such as Intermediate Fabrics \cite{IntermediateFabric2010}  to meet the applications' timing requirements.

\subsection{Software Stack}
\label{subsec3.2:Software Stack}
The software stack refers to an OS, software applications, and frameworks running on a host CPU for the purpose of FPGA virtualization. A host CPU can be either on-chip, local, or remote, depending on the hardware stack. The abstraction of vFPGA is generally made available to users in the software stack. The software stack allows users to develop and deploy applications onto vFPGAs and manage vFPGA resources easily without specific knowledge of the underlying hardware. Specifically, the software stack provides users with software libraries and APIs to perform the communication between a host and an FPGA computing device, the provisioning and management of applications and physical FPGA resources. Table \ref{tab:table-1} shows a list of the software stack used in the existing FPGA virtualization work in three categories: OS, host application, and software framework.

\textbf{OS:} Since the idea of FPGA virtualization is primarily adopted from OS, the most common software stack of FPGA virtualization is an OS (\textit{e.g.}, Windows and Linux), often supporting multiple processes and threads \cite{VERCloud2017,VirtualRuntime2016} for the management of multiple users and FPGA devices. In FPGA virtualization, both compile- and run-time management of FPGA resources are important factors to be considered. If the virtualization system is designed for a cloud or edge computing environment where real-time streaming data is used, OSs with real-time processing capability (\textit{e.g.}, real-time operating system (RTOS) \cite{Ker-ONE2016}) is preferable. There are specialized OSs particularly designed for reconfigurable computing and FPGAs. For example, ReconOS \cite{ReconOS2014} is designed to support both software and hardware threads, as well as allow interaction between the threads in a reconfigurable computing system. Leap FPGA OS \cite{LeapFPGA2014}) and Operating System for Reconfigurable Systems (OS4RS) \cite{OS4RS2009} are FPGA virtualization OS for compile-time management of FPGA resources. RACOS \cite{RACOS2017} provides a simple and efficient interface for multiple user applications to access single and multi-threaded accelerators transparently. AmorphOS’s OS \cite{khawaja2018sharing}, integrated as a user-mode library on a host CPU, provides system calls to manage Morphlets (\textit{i.e.}, vFPGAs) and enables the communication between host processes and Morphlets.

\textbf{Host Application:} Application running on host CPUs are commonly written in C/C++ or OpenCL. OpenCL provides the development language for device kernels as well as the programming API for host applications, which makes it a good choice for FPGA virtualization \cite{FPGAs-Cloud2017,ResourceElastic2018}. The programming API can be used to deploy, manage, and communicate with FPGA kernels. Vendors may provide SDKs for OpenCL (\textit{e.g.}, Intel FPGA SDK for OpenCL \cite{IntelSDKOpenCL:2020}) that provides the API for a host OpenCL application to manage the execution of FPGA kernels. One important task of the host application is the management of communication between the host and the FPGA device. In \cite{VirtualRC2012}, a software-middleware API, written in C++, is provided to enable the portability of application software code (\textit{i.e.}, host code) on various physical platforms. The software-middleware is a virtualization layer between the application software and the platform API. The middleware API translates the virtual platform communication routines for the virtual components into native API calls to the physical platform in \cite{pvFPGA2013}, the concept of virtual machine monitor (VMM) \cite{barham2003xen} used in OS virtualization is adopted in FPGA virtualization for the communication between a user and an FPGA device. They have modified the existing inter-domain (\textit{i.e.}, driver domain, Dom0, and unprivileged domain, DomU) communication in Xen VMM using shared memory. In this way, multiple user processes can simultaneously access a shared FPGA with reduced overhead.

\textbf{Software Framework:} OpenStack \cite{OpenStack:2020} is an open-source platform for controlling compute, storage, and networking resources in cloud computing. Some of the FPGA virtualization systems developed for cloud and data center utilize the OpenStack framework \cite{AcceleratorPool2014,VFR2014,HSDC2015}). In \cite{AcceleratorPool2014}, new modules are added to OpenStack compute nodes (\textit{i.e.}, a physical machine composing of CPUs, memory, disks, and networking) in different layers (e.g., hardware, hypervisor, library, and layers) to support FPGA resource abstraction and sharing.  In \cite{HSDC2015}, an accelerator service is introduced in OpenStack to support network-attached standalone FPGAs. The FPGAs have software-defined-network-enabled modules that can connect to FPGA plugins in the accelerator service. Byma et al. \cite{VFR2014} uses a Smart Application on Virtual Infrastructure (SAVI) test framework with OpenStack to integrate FPGAs to cloud and manage them like conventional VMs. In this work, the authors proposed to partition the FPGA into multiple PRRs and abstract each PRR as a Virtualized FPGA Resource (VFR). An agent application with a device driver is developed to manage the VFRs. The SAVI testbed controller in OpenStack provides the APIs for connecting the VFR agent with the driver, enabling the remote management of FPGA resources from a cloud server. Fahmy et al. \cite{VFACloud2015} extends an existing FPGA test platform called DyRACT \cite{DyRACT2014} to support multiple independent FPGA accelerators. A software framework and a hypervisor is implemented in DyRACT to connect with the communication interface in the static region of an FPGA. FPGAVirt \cite{FPGAVirt2018, VirtioHWSB2018} leverages a communication framework named Virtio-vsock \cite{Virtio-vsock2015} to develop their software virtualization framework, in which Virtio-vsock provides an I/O virtualization framework and a client for the communication between VMs and FPGAs.

\subsection{Overlay Architecture}
\label{subsec3.3:Overlay}
\label{subsec:overlay architecture}
An FPGA overlay, also known as intermediate fabric \cite{IntermediateFabric2010}, is a virtual reconfigurable architecture implemented on top of a physical FPGA fabric for providing a higher abstraction level \cite{So2016} of FPGA resources. An overlay architecture serves as an intermediate layer between a user application and a physical FPGA. The physical architecture of an FPGA can vary significantly across different device families or vendors. Overlay architectures can bridge that gap by providing a uniform architecture on top of the physical FPGA. The overlay's application software is portable on any device which supports the targeted overlay architecture. This concept is analogous to Java Virtual Machine \cite{JVM2014} that allows the same byte-code to be executed on any Java-supported machines. Overlay architectures provide a much higher level of hardware abstraction for application mapping, which significantly reduces the compilation time and provides improved portability of application codes across different FPGA device families and vendors. Overlay architectures also provide application developers with better software design flexibility as they can target an abstract computing architectures with a known instruction set. Furthermore, portability in FPGA resource management can be achieved using overlay architectures.  Resource management techniques can vary significantly depending on the FPGA architecture and the implementation stack where the resource manager is implemented (Section \ref{subsec4.3: Resource Management}). If the resource manager is implemented in the overlay stack, the same overlay architecture, therefore same techniques can be applied to manage PRR of different FPGA devices.  However, these benefits come at the cost of reduced performance and less efficient utilization of FPGA resources. 

\textbf{Configuration:} An FPGA overlay can be either spatially configured (SC) or time-multiplexed (TM), depending on its run-time configurability. If an FPGA overlay has functional units with fixed assigned tasks, it is referred to as an SC overlay. If an FPGA overlay can change the operation of its functional units on a cycle-by-cycle basis, the overlay is referred to as a TM overlay. The interconnection between functional units in SC and TM overlays can be fixed or reconfigurable at run time, which can be structured as a Network-on-Chip (NoC). A previous survey paper \cite{Li:2019:TFO:3339837.3339861} provides a comprehensive survey of TM overlays. 

\textbf{Granularity:} According to the granularity (at which hardware can be programmed) of an overlay architecture, SC and TM overlays can be further categorized into fine-grained and coarse-grained overlays \cite{HeterogeneousCGRA2011}. Fine-grained overlays can operate at the bit-level just like physical FPGAs but provide programmability and configurability at a higher level of abstraction than physical FPGAs. Firm-core virtual FPGA \cite{Firm-coreJIT2005} is an early work in the area of fine-grained overlays, where a synthesizable VHDL model of a target FPGA fabric is developed. Two types of switch matrices developed using tri-state buffers and multiplexers provide the programmable interconnect between the custom CLB interfaces. Even though the overlay has huge hardware overhead, it can be used in applications where portability is more important than resource utilization. ZUMA \cite{Brant2012ZUMAAO} also provides bitstream portability across FPGAs of different vendors with a more optimized implementation than Firm-core virtual FPGA. ZUMA configures LUTRAMs to use them as the building block for configurable LUTS and routing multiplexers. A configuration controller is implemented to reprogram the LUTRAMs connected in a crossbar network. A previous survey paper in \cite{Jain2017} presents a comprehensive survey about coarse-grained overlays. Coarse-grained overlays are implemented as coarse-grained reconfigurable arrays (CGRAs) \cite{CGRA2001} or processors. In a CGRA, arrays of PEs are connected via a 2D network. CGRAs can adopt different interconnect topologies (\textit{e.g.}, island style, nearest neighbor, and NoC). The 2D mesh structures in the

\begin{landscape}
\centering
\begin{table}[b]
\scriptsize
\caption{\label{tab:table-3} Summary of FPGA Virtualization Techniques for Abstraction, Multi-tenancy, Resource Management, and Isolation}

\begin{threeparttable}
\tiny

\renewcommand{\arraystretch}{2}
\begin{tabular}{|c|p{2cm}|c|c|p{0.7cm}|c|p{0.7cm}|c|c|}

\hline
\multirow{2}{*}{\textbf{Work}} & \multicolumn{2}{c|}{\textbf{Abstraction}} & \multicolumn{2}{c|}{\textbf{Multi-tenancy}} & \multicolumn{2}{c|}{\textbf{Resource Management}} & \multicolumn{2}{c|}{\textbf{Isolation}} \\ \cline{2-9} 
 & \textbf{Technique} & \textbf{IS} & \textbf{User management} & \textbf{MT} & \textbf{Technique} & \textbf{IS} & \textbf{Technique} & \textbf{IS} \\ \hline \hline
 
Virtual \cite{VirtualFPGA1998,GeneralVirtualization1998} & N/A & N/A & N/A & N/A & Partitioning, Segmentation, Overlaying & SW, HW & N/A & N/A \\ \hline
Firm-core JIT \cite{Firm-coreJIT2005} & N/A & N/A & N/A & N/A & Firm Core intermediate fabric & OL & N/A & N/A \\ \hline
Cray HPRC \cite{CrayHPRC2008} & N/A & N/A & N/A & SM & Virtualization Manager, Virtual Memory Space, Message Queue & SW & N/A & N/A \\ \hline
OS4RS \cite{OS4RS2009} & Unified communication interface between HW and SW & SW, HW & N/A & SM, TM & virtual hardware in OS and unified communication mechanism w/ HW & SW, HW & N/A & N/A \\ \hline
NoC \cite{HomogeneousNoC2010} & N/A & N/A & N/A & SM, TM & OS with run-time reconfiguration and virtual FPGA page table management & SW & N/A & N/A \\ \hline
Fabric\cite{IntermediateFabric2010} & Virtual abstract architecture using intermediate fabric & OL & N/A & TM & N/A & N/A & N/A & N/A \\ \hline
VirtualRC\cite{VirtualRC2012} & N/A & N/A & N/A & N/A & virtual FPGA platform and software-middleware API & SW & N/A & N/A \\ \hline
HPRC \cite{HPRCVCM2012} & N/A & N/A & N/A & TM & Virtualization Manager, Virtual Memory Space, Message Queue & SW & N/A & N/A \\ \hline
pvFPGA\cite{pvFPGA2013} & N/A & N/A & N/A & TM & Frontend and Backend Driver, Device Driver for FPGA accelerator with Xen VMM & SW & N/A & N/A \\ \hline
VDI \cite{HWSWVDI2013} & N/A & N/A & N/A & SM & HW Task Manager, HW Control Library & SW & N/A & N/A \\ \hline
ReconZUMA\cite{ReconOSZUMA2014} & ReconOS with ZUMA Overlay & SW, OL & N/A & N/A & N/A & N/A & N/A & N/A \\ \hline
ACCL\cite{AcceleratorPool2014} & N/A & N/A & OpenStack controller & SM, TM & Hypervisor layer in SW and service layer on FPGA & SW, HW & N/A & N/A \\ \hline
VFR \cite{VFR2014} & N/A & N/A & OpenStack controller & SM, TM & PRRs managed by OpenStack & SW, HW & N/A & N/A \\ \hline
HSDC \cite{HSDC2015} & N/A & N/A & OpenStack controller & SM & Network Manager and Accelerator Service Extension for OpenStack & SW, HW & N/A & N/A \\ \hline
VFA Cloud \cite{VFACloud2015} & N/A & N/A & Software running on client machine & SM, TM & Hypervisor on cloud virtualization layer & SW & N/A & N/A \\ \hline
RC3E \cite{RC3E2016} & N/A & N/A & Reconfig. Common Cloud Computing Environment & SM, TM & Hypervisor on cloud for integration and management of virtual hardware & SW, HW & N/A & N/A \\ \hline
RRaaS\cite{RRaaS2016} & N/A & N/A & Application for user request management & SM, TM & controller/Hypervisor, NoC architecture of reconfigurable regions & SW, HW & N/A & N/A \\ \hline
Virt. Runtime \cite{VirtualRuntime2016} & N/A & N/A & User threads and thread manager & SM, TM & run-time manager on on-board processor, FPGA threads & HW & N/A & N/A \\ \hline
Ker-ONE\cite{Ker-ONE2016} & N/A & N/A & N/A & SM & Hypervisor on ARM processor and PRR Monitor on FPGA & SW, HW & FI of VMs and DPRs using Hypervisor on ARM processor & SW \\ \hline
RC2F \cite{RC2F2017} & N/A & N/A & N/M & SM, TM & Hypervisor on cloud for integration and management of virtual hardware & SW, HW & N/A & N/A \\ \hline
VER Cloud\cite{VERCloud2017} & N/A & N/A & Multi-threading on host CPU & SM, TM & run-time manager on soft processor & OL & N/A & N/A \\ \hline
Feniks \cite{Feniks2017} & Identical virtual IO interface & SW, HW & N/M & SM & OS on FPGA, host agent connected to centralized controllers in cloud & SW, HW & PI by separation of application and OS & HW \\ \hline
NFV \cite{NFVManager2017} & N/A & N/A & Network requests on host OS & SM, TM & Context Manager implemented as MCU on soft/SoC processor & HW, MW & N/A & N/A \\ \hline
FPGAs-Cloud \cite{FPGAs-Cloud2017} & N/A & N/A & VM and Hypervisor on host & SM & Hypervisor in static region of FPGA & HW & N/A & N/A \\ \hline
Virt. Security\cite{VirtualizationSecurity2017} & N/A & N/A & N/A & N/A & N/A & N/A & FI using unique Overlay & OL \\ \hline
Cost of Virt \cite{CostofVirtualization2017} & N/A & N/A & N/A & N/A & Avalon interconnect for managing PRRs & HW & N/A & N/A \\ \hline
FPGAVirt \cite{VirtioHWSB2018, FPGAVirt2018} & NoC Overlay architecture & OL & Virtio Vsock client running on guest OS & SM, TM & FPGA Management Service, Mapping Table, NoC Overlay architecture & SW, OL & FI using Hardware Sandbox & HW \\ \hline
hCODE \cite{hCODE2018} & N/A & N/A & Scheduler on host & SM & Multi-channel PCIe module to manage PRRs & HW & N/A & N/A \\ \hline
Elastic \cite{ResourceElastic2018} & N/A & N/A & N/A & SM, TM & Run-time resource manager for virtualizing resource footprint of OpenCL kernels & SW & N/A & N/A \\ \hline
AmorphOS \cite{khawaja2018sharing} & Morphlet which encapsulates user FPGA logic & SW, HW & N/A & SM, TM & zone manager on host CPU & SW & FI and PI using resource allocation policy and hw arbiter & SW, HW \\ \hline
ViTAL\cite{zha2020virtualizing} & Layer between physical FPGAs and compilation layer & SW, HW & N/A & SM, TM & Hypervisor and system controller & SW & FI using runtime management policy & SW \\ \hline
Shared Mem\cite{SharedmemHypervisor2020} & N/A & N/A & VM and Hypervisor on host & SM, TM & Hypervisor and hardware monitor for managing accelerators & SW, HW & FI between accelerators using page table slicing & SW, HW \\ \hline
\end{tabular}%
\begin{tablenotes}
\begin{tiny}
\item[\tnote{\textdagger}] 
N/A = Not Applicable; N/M = Not Mentioned; IS = Implementation Stack; MT = Multiplexing Technique; SM = Spatial Multiplexing; TM = Time Multiplexing; FI = Functional Isolation; PI = Performance Isolation;
\end{tiny}
\end{tablenotes}
\end{threeparttable}
\end{table}
\end{landscape}

\noindent island style and nearest neighbor typologies have similarities to the interconnect on FPGAs. These interconnects support direct communication between adjacent PEs. However, the communication flexibility comes at the cost of additional FPGA resource usage.  In the NoC topology, the PEs are connected in a network and communicates via routers. NoC-based CGRA architectures are becoming popular in FPGA virtualization due to the flexible communication between PEs at run time.

Overlay architectures have been discussed in the literature for a long time, even before the concept of FPGA virtualization. The primary purpose of overlay architecture has been to improve design productivity and reduce the compilation time of FPGA design. FPGA overlays, especially CGRA-based overlays, have recently caught a lot of attention for being used for FPGA virtualization. In a recent work \cite{FPGAVirt2018}, authors present an NoC-based overlay \cite{Flexitask2018} architecture that offers flexible placement of hardware tasks with high throughput communication between PEs. The architecture has a torus network equipped with data packet communication and high-level C++ library for accessing overlay resources. Adjacent PEs form a sub-network, in which they communicate directly with each other, and the inter-subnetwork communication takes place via routers. 

\begin{figure}[]
  \includegraphics[width=3.3in]{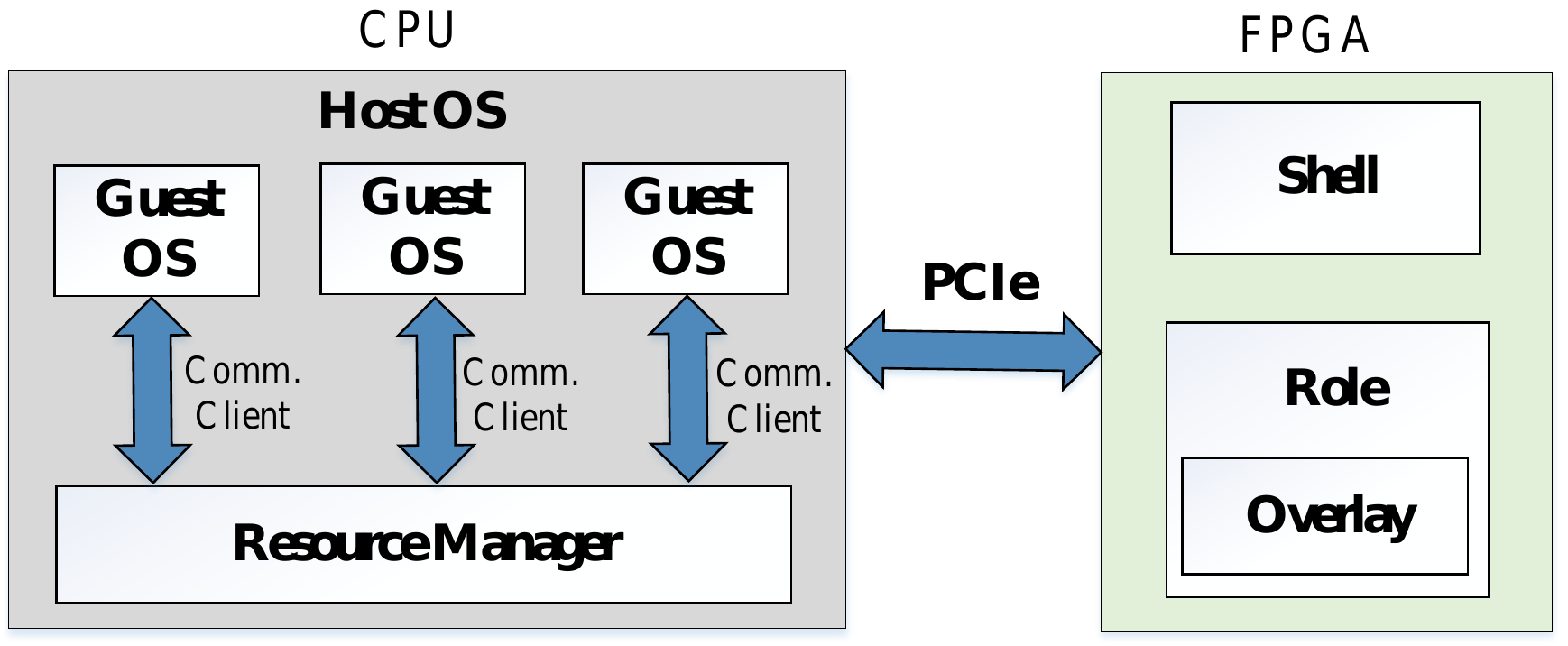}
  \caption{System architecture for an exemplar FPGA virtualization system}
  \label{fig:sysarch}

\end{figure}%

Figure \ref{fig:sysarch} shows a block diagram of an example FPGA virtualization system, generalized from the system in \cite{FPGAVirt2018}. Description of different stacks of the system architecture is shown in Table \ref{tab:table-sysarch}.

\begin{table}[]
\centering
\caption{\label{tab:table-sysarch} Description of the system architecture stacks of the example FPGA virtualization system in Figure 2}

\begin{tabular}{|l|p{7cm}|}
\hline
\multicolumn{1}{|c|}{\textbf{Stack}} & \multicolumn{1}{c|}{\textbf{Description}} \\ \hline
\multirow{3}{*}{\textbf{HW}} & \begin{tabular}[c]{@{}l@{}}The host interface is a local host interface where \\ the CPU is connected to the FPGA via PCIe.\end{tabular} \\ \cline{2-2} 
 & \begin{tabular}[c]{@{}l@{}}Shell: Static region that contains the board support\\ package (BSP).\end{tabular} \\ \cline{2-2} 
 & \begin{tabular}[c]{@{}l@{}}Role is the reconfigurable region of the FPGA. Here, an \\ overlay architecture is implemented on the\\ reconfigurable region.\end{tabular} \\ \hline
\multirow{3}{*}{\textbf{SW}} & \begin{tabular}[c]{@{}l@{}}A host OS in the CPU hosting multiple virtual \\ machines with guest OSs. Users in guest OS can \\ make requests to use the FPGA.\end{tabular} \\ \cline{2-2} 
 & \begin{tabular}[c]{@{}l@{}}A resource management software that receives \\ requests from guest OS and communicates with \\ FPGA to fulfill the requests. It also keeps track \\ of the available FPGA resources.\end{tabular} \\ \cline{2-2} 
 & \begin{tabular}[c]{@{}l@{}}A communication client that manages communication \\ between guest OS and resource manager.\end{tabular} \\ \hline
\textbf{OL} & \begin{tabular}[c]{@{}l@{}}A CGRA based overlay that creates an abstraction \\ of the reconfigurable region in physical FPGA.\end{tabular} \\ \hline
\end{tabular}%

\end{table}

\section{Objectives of FPGA Virtualization}
\label{sec:virtualization objectives}
As discusses in section \ref{sec:background}, the primary objectives of FPGA virtualization are abstraction, multi-tenancy, resource management, and isolation. In this section, we summarize how these four key objectives are achieved in the literature. 

\subsection{Abstraction}
\label{subsec4.1:Abstraction}
The first objective of FPGA virtualization aims to make FPGA more usable by creating abstractions of the physical hardware to hide the hardware specifics from application developers and the OS as well as provide applications with simple and familiar interfaces to access the virtual resources. Table \ref{tab:table-3} lists a summary of FPGA virtualization techniques for creating abstraction, which can be categorized into overlay architecture and OS-based approaches. 

As discussed in Section \ref{subsec3.3:Overlay}, overlay architecture provides a known-instruction-set architecture that can be targeted by software developers to create and run applications for FPGAs. The usability benefit comes with a penalty of performance drop and less efficient FPGA resource usage. This issue can be mitigated by using a database of overlay architectures and selecting the optimized one for each application \cite{QuickDough2015}. Overlay architecture also benefits usability by significantly reducing the compilation time of application codes and providing homogeneity in abstraction across the FPGA devices from different vendors. For example, the same overlay architecture implemented on Intel and Xilinx FPGAs provides exactly the same abstraction even though the underlying hardware architecture is different. 

OSs used in FPGA virtualization provide software interfaces for easier deployment and access to FPGA applications. Some examples of work in this area are Feniks OS \cite{Feniks2017}, ReconOS \cite{ReconOS2014}, RACOS \cite{RACOS2017}, OS4RS \cite{OS4RS2009}, and AmorphOS \cite{khawaja2018sharing}. Feniks OS extends the shell region of an FPGA to implement the OS, which provides an abstracted interface to developers for application management. In OS4RS, the concept of a device node is proposed where multiple applications on an FPGA can be accessed using a single node. RACOS provides users with a simple interface to load/unload accelerators and presents the I/O transparently to the users.AmorphOS’s OS interface exposes APIs to load and deplete Morphlets (\textit{i.e.}, vFPGAs), and read and write data over the transport layer to the physical FPGAs configured with Morphlets.
The abstraction created for FPGA virtualization not only facilitates application development but also enables better application access. In \cite{AcceleratorPool2014}, an accelerator marketplace concept is presented where users can buy and use accelerator applications. The marketplace is implemented on top of an FPGA virtualization layer integrated into OpenStack. The underlying virtualization layer hides the complexity of hardware to make it extremely easy to develop, manage and use FPGA applications.

\subsection{Multi-Tenancy}
\label{subsec4.2: Multi-Tenancy}
The second objective of FPGA virtualization aims to share the same FPGA resources among multiple users and applications. Since in the FPGA virtualization infrastructure, it is necessary to support multiple applications, we narrow our discussion in this subsection to multiple users. Therefore, single/multiple tenant refers to if an FPGA virtualization system supports resource sharing across multiple users or not.

For multi-tenancy support, it is required to have either spatial or temporal multiplexing. In spatial multiplexing, by utilizing PR techniques, an FPGA device is partitioned into multiple PRRs such that each PRR can be reconfigured independently for application mapping without affecting other regions. The reconfiguration can be done either statically or dynamically at run time. In temporal multiplexing, there is only one PRR, which is reconfigured over time, and different applications are allocated in different time intervals. The reconfiguration time of the PRR should be reduced as much as possible to maintain efficient resource sharing. 

The user management components are generally implemented in the software stack. Table \ref{tab:table-3} shows a list of user management techniques used in the literature. The most common methods in the literature to support multiple users include utilizing Openstack Control Node \cite{AcceleratorPool2014,VFR2014,HSDC2015}, Virtio-vsock client running on a guest OS \cite{VirtioHWSB2018,FPGAVirt2018}, and multi-threading on a host CPU \cite{VERCloud2017,VirtualRuntime2016}.

In \cite{VFR2014}, the OpenStack scheduler selects a resource and contacts its associated agent, which is a separate process beside the hypervisor, upon a request for a vFPGA from a user. When OpenStack requests the PRR from the agent, it commands the hypervisor to boot a VM with the user-defined OS image and parameters. In \cite{AcceleratorPool2014}, a control node manages requests from multiple users and launches VMs on physical machines. Subsequently, different users access their associated VMs and deploy their applications. In \cite{HSDC2015}, when a user requests for a vFPGA, the accelerator scheduler searches in a resource pool to find a PRR matching the user request. If successful, a user ID and IP address will be configured for the vFPGA. Subsequently, the vFPGA ID, IP address, and required files to generate a bitstream for the user application will be returned to the user.

In \cite{VirtioHWSB2018} and \cite{FPGAVirt2018}, utilizing an overlay architecture enables assigning multiple vFPGAs on a single or several FPGAs. Each user accesses a guest OS installed on a VM. Upon the user request, the corresponding VM requests FPGA resources, and the hypervisor looks for an available PRR. If available, memory space is allocated to the vFPGA for storing data.

In \cite{VirtualRuntime2016} and \cite{VERCloud2017}, a run-time manager that provides hardware-assisted memory virtualization and memory protection allows multiple users to simultaneously execute their applications on an FPGA device. A user thread on a host CPU sends the specifications of vFPGAs and codes to be run on a local processor to the manager thread, which will be deployed to an FPGA. On the FPGA side, a run-time manager allocates the resources required by the FPGA application, creates an FPGA user thread with the codes for the local processor, instantiates a vFPGA, and notifies the host when the FPGA application terminates. The host user thread can then retrieve the output data from the FPGA and either sends new data to be processed or deallocates the FPGA application.

\subsection{Resource Management}
\label{subsec4.3: Resource Management}
The third objective of FPGA virtualization aims to facilitate the transparent provisioning and management of FPGA resources for workload balancing and fault tolerance. The management of FPGA resources refers to the management of the PRRs. For a virtualization system with a single FPGA, resource management primarily involves configuring the PRRs by allocation and deallocation of bitstreams. In a virtualization system where multiple FPGAs are connected via a network, routing of information among multiple FPGAs, configuring the PRRs of multiple FPGAs with bitstreams, and the scheduling of accelerators can be labeled as resource management tasks. Table \ref{tab:table-3} lists a summary of FPGA virtualization techniques for resource management, which are discussed below according to which stack (software, hardware, overlay) in the system architecture they are implemented in.   

\subsubsection{\label{re-sw}Resource Management in the Software Stack }
The primary ideas of the resource management approaches implemented in the software stack are borrowed from OS, CPU, and memory virtualization. Partitioning methods in memory systems (\textit{e.g.}, paging and segmentation) are used to divide and keep track of the FPGA resource utilization. The OS in \cite{HomogeneousNoC2010} used virtual FPGA page tables and a reconfiguration manager to manage configurations of identical PRRs connected with NoC-based interconnects. While the virtual page table keeps track of the resource utilization of the PRRs, the run-time reconfiguration manager swaps configurations in and out of the PRRs. Other than partitioning and page tables, the concept of a hypervisor in a virtual machine has also been widely used in FPGA virtualization for resource management. A hypervisor is generally part of the host machine and used for the management of and communication with hardware resources on an FPGA. Fahmy et al. \cite{VFACloud2015} implements a hypervisor as a part of the server-side software stack in a cloud-based architecture. The hypervisor communicates with the PRRs via a communication interface implemented in the FPGA static region. The hypervisor is responsible for maintaining a list of PRRs, configuring vFPGAs by selecting optimal PRRs from that list, as well as allocating vFPGAs to users. In \cite{SharedmemHypervisor2020}, the authors propose a hypervisor named Optimus for a shared-memory FPGA platform. The hypervisor provides scheduling of VMs on a pre-configured FPGA with temporal multiplexing to share a single accelerator with multiple VMs as well as spatial multiplexing to share the FPGA among multiple accelerators. Optimus receives requests from host applications as interrupts and communicates with accelerators using a hardware monitor implemented in the shell. Knodel et al. present a cloud-based hypervisor called RC3E in \cite{RC3E2016}, which integrates and manages FPGA-based hardware accelerators in the cloud. The resource manager, termed as FPGA device manager in RC3E, configures vFPGA using bitfiles from a database and monitors the accelerators accessed by the user virtual machines. The Reconfigurable Common Computing Framework (RC2F) used in RC3E was extended by the authors in \cite{RC2F2017}. In this work, the authors present one physical FPGA as multiple vFPGAs and manage the vFPGAs using their custom FPGA hypervisor. The FPGA hypervisor manages the states of the vFPGAs, as well as reconfigures them using the internal configuration access port (ICAP). In SoC-based FPGAs, the hypervisor can run in an OS running on an on-chip CPU. For example, in \cite{Ker-ONE2016}, a hypervisor implemented on an ARM processor detects the requests from different VMs and programs PRRs dynamically. Some of the prior work utilizes existing software frameworks such as OpenStack or OpenCL run-time to manage FPGA resources, which is discussed in section \ref{subsec3.2:Software Stack}. 

\subsubsection{Resource Management in the Overlay Stack }
Similar to the SW stack, the resource managers in the overlay stack are also implemented in software. However, they are implemented inside the FPGA, therefore, are tightly coupled with the PRRs in FPGA. The resource managers in the SW stack discussed in section \ref{re-sw} are implemented in the host outside of the FPGA and are loosely coupled with the PRRs they manage. In \cite{VERCloud2017}, a soft-processor-based overlay runs a run-time manager or hypervisor for FPGA resource management in a similar way as to how an OS on a host machine manages the hardware resources. The run-time manager can handle service requests as interrupts from accelerators as well as from the host. The interrupt from accelerators is received using a custom-designed message box module, while the interrupt from the host is received using PCIe. In \cite{NFVManager2017}, the resource manager is implemented as a microcontroller unit (MCU) running on a soft-processor-based overlay, which acts as a scheduler and context manager of accelerators. Context management within the same accelerator is handled using job queues, while multiple accelerators are managed using a job scheduler. Other than processor-based overlays, some of the CGRA-based overlays are specifically designed for resource management. For example, NoC-structured CGRA-based overlay \cite{Flexitask2018} architectures are designed for better placement of applications into PRRs. The overlay abstracts the PRRs as connected PEs in a 2D Torus topology with high throughput communication between the PEs using routers. Placement of applications in adjacent PEs can be done directly, whereas the placement to distant PEs is done using the routers. 

\subsubsection{Resource Management in the Hardware Stack }
In the hardware stack, a resource manager is generally implemented as part of the static region. In this case, a resource manager is in charge of the communication with a host machine using PCIe and configuration access ports (\textit{e.g.} ICAP and PCAP in Xilinx FPGA \cite{PRUserGuide:2018}). In \cite{hCODE2018}, the authors built a multi-channel shell with multiple independent PCIe channels for managing multiple accelerators. A command-line tool is used to load bitstreams from a repository and send over the PCIe channels to program the vFPGAs. In \cite{Ker-ONE2016}, a virtual device manager on a host soft processor sends configuration requests to a separate resource manager in the static region. The resource manager downloads bitstreams using the PCAP interface and programs the PRRs via interconnects between the resource manager and PRRs.  Similar to the software stack, a hypervisor can be implemented in the hardware stack. Kidane et al. \cite{RRaaS2016} propose a hypervisor implemented in the static region of an FPGA to manage vFPGA resources. The hypervisor is responsible for the selection of bitstream from a library and the placement of the bitstream to an appropriate vFPGA based on the size of the design. In \cite{AcceleratorPool2014}, the resource manager is implemented in the static region that provides a communication interface with a hypervisor in the software layer as well as an interface for managing PRRs. The resource manager fetches requests from the software stack from a job queue and programs the PRRs. It also uses a direct memory access (DMA) engine to context-switch and to manage data to/from accelerators. Tarafder et al. \cite{FPGAs-Cloud2017} uses the Xilinx SDAccel platform \cite{SDAccel:2016} as a hypervisor implemented on the static region and provides the interfaces to memory, PCIe, and Ethernet. The hypervisor can program the PRRs once it receives configuration requests from the OpenCL API via PCIe. Custom state machines or hardware controllers can also be used for resource management. Custom-designed hardware controllers can leverage the DMA transfer of pre-stored bitstreams in off-chip memory for faster configuration. In DyRACT \cite{DyRACT2014}, resource management is done by two custom state machines named the configuration state machine and the ICAP state machine. The configuration state machine receives configuration requests from PCIe and writes the bitstream in memory, while the ICAP state machine reads the bitstream from memory and programs the FPGA. 

FPGA resource management approaches implemented in the software stack benefit from the convenience of software development and reuse of existing software frameworks. However, the communication between a host CPU and an FPGA can become a bottleneck and hurt the management performance. As different FPGA vendors have different techniques for managing PRRs in their FPGA devices, handling the management of PRRs with overlay architectures has the benefit of allowing the resource management approaches to be compatible with a variety of FPGA devices across different vendors. However, such a benefit comes with penalties on performance and resource utilization efficiency. Differently, implementing resource management utilities in the hardware stack increases the management performance but loses the compatibility as well as reduces the available FPGA resources for application mapping.

\subsection{Isolation}
\label{subsec4.4:Isolation}
The fourth objective of FPGA virtualization aims to provide strict performance and data isolation among different users and applications to ensure data security and system resilience. Table \ref{tab:table-3} summarizes the list of papers that addresses the isolation objective in FPGA virtualization.

If the FPGA virtualization system supports multi-tenancy, each tenant in the system will have shared access to the available hardware resources on the FPGA, \textit{e.g.} one or multiple PRRs, a set of I/O resources, and on-chip or off-chip memory resources. Therefore, it is essential to make sure each tenant only has strict access to its own resources at a specific time, and they are not able to access or manipulate data nor interrupt the execution of accelerators of other tenants. 

From the perspective of the hardware stack, at least three types of isolation are required for FPGA virtualization systems: functional isolation, performance isolation, and fault isolation. 

Functional isolation ensures that different vFPGAs can be reconfigured and managed independently without affecting the functionality and operation of the others. The DPR capability of modern FPGA supports the independent reconfiguration of a PRR at run time while other PRRs are active, providing the fundamental support for realizing functional isolation. Functional isolation also ensures that each user or application only has access to its assigned hardware resources. In FPGA virtualization systems, memory and I/O resources are often shared among multiple users and/or applications in a time-multiplexed fashion. Therefore, safe memory and I/O access need to be ensured to guarantee the integrity of functionality and avoid data conflicts. Existing locking mechanisms such as mutex and semaphore \cite{VERCloud2017} can be used to ensure safe memory access. Moreover, when multiple applications share an FPGA, strict isolation among their I/O channels is vital to ensure the correct transmission of data. Some FPGA vendors (\textit{e.g.}, Xilinx) have proposed the isolation design flow \cite{IsolationDesignFlow:2020} to ensure the independent execution of multiple applications on the same FPGA device. In \cite{SharedmemHypervisor2020}, the hypervisor designed by the authors provides memory and I/O isolation using a technique called page table slicing. In page table slicing, each guest application and their accelerators use their guest virtual address to interact with the IO address space of the DRAM on the FPGA device. The hypervisor partitions the IO address space among accelerators so that each accelerator has a unique memory and IO address space that guest applications can access. The hypervisor maintains an offset table for address translation between the guest virtual address and IO address space. The FPGA virtualization frameworks proposed in \cite{FPGAVirt2018, VirtioHWSB2018} utilize Hardware Sandbox (HWSB) for the isolation of vFPGAs based on their overlay architecture \cite{Flexitask2018}. The HWSB adds the identifier of a target vFPGA to a data packet while communication over the overlay's interconnection routers among different vFPGAs belonging to the same VM, which prevents unauthorized access of data \cite{VirtioHWSB2018}. The work in \cite{VirtualizationSecurity2017} utilizes virtual architectures (i.e., overlays) to safeguard FPGAs from bitstream tampering. In this work, an application-specialized overlay is created using an overlay generator. Then, for each vFPGA that deploys the same application as other vFPGAs, the generated overlay bitfile is modified to create a unique overlay. This process is called uniquification and improves bitstream security. In \cite{Ker-ONE2016}, the functional isolation between VMs and PRRs is ensured by implementing a custom VMM on an ARM processor. Each VM has its dedicated software tasks and isolated virtual resources managed by the VMM.

Performance isolation refers to the capability of limiting the temporal interference among multiple vFPGAs running on the same physical FPGA so that the event of a vFPGA will have little impact on the performance of other vFPGAs. For example, in cloud-based FPGA virtualization, when different user requests are handled dynamically by different vFPGAs at run time, the performance of each vFPGA should stay stable as long as the peak processing capability of the physical FPGA is not reached. Feniks OS \cite{Feniks2017} provides performance isolation in the role by using PR techniques, as PR disables interconnections and logic elements on the boundary of PRRs. Therefore, adjacent PRRs are physically unable to affect each other. In addition, this work separates the OS from the application region by implementing the OS in the shell.

Fault isolation refers to the capability of limiting the negative impacts of software faults or hardware failures on the operation of vFPGAs for system resilience. The hardwired PCIe controllers and DPR capability of modern FPGAs inherently provide the hardware-level support for fault isolation. In addition, system-level fault recovery and fallback mechanisms are required to fast-recover vFPGA services to further improve the system resilience. In \cite{catapult2014}, the authors introduced a protocol to ensure fault isolation. The protocol can reconfigure groups of FPGAs or remap services robustly, re-maps the FPGAs to recover from failure, and report a vector of errors to their management software to diagnose problems.

Integrating FPGAs into a multi-tenant environment makes them vulnerable to remote software-based power side-channel attacks \cite{krautter2019active, yazdanshenas2019costs}. To tackle the issue, an on-chip power monitor can be programmed on a region dedicated to an attacker on a shared FPGA using ring oscillators \cite{zhao2018fpga}, and the power monitor can observe the power consumption of other regions on the FPGA. The observed power consumption could reveal sensitive information such as bit value in the RSA crypto engine. Moreover, the malicious application could cause delay faults for the victim region by disturbing the power network through aggressive consumption of power. Unfortunately, there has not been much work aiming to address such security issues. To adopt the multi-tenancy techniques in FPGA virtualization systems discussed in section \ref{subsec4.2: Multi-Tenancy}, these security issues need to be addressed carefully in future research.

Although isolation is an important topic for the practical application of FPGA virtualization systems, especially in the context of multi-tenancy cloud and edge computing, there has not been much work focusing on this topic in the existing literature. Future research in this area, especially the fault isolation in FPGA virtualization, is critically needed. 

\section{Conclusion and Future Trends}
\label{sec:conclusion}
This paper provides a survey on the system architectures and various techniques for FPGA virtualization in the context of cloud and edge computing, which is intended to facilitate future researchers to efficiently learn about FPGA virtualization research. 

From a careful review of the existing literature, we identified two research topics in FPGA virtualization that need extra attention in future research. First, we find that the hardware boundary of a vFPGA is limited to a PRR in a single FPGA in most of the existing work, which has limited the scope of FPGA virtualization. Ideally, FPGA virtualization should completely break the hardware boundary of FPGAs such that a physical FPGA can be used as multiple vFPGAs, and multiple physical FPGAs (connected on-package, on-board, or via network) can also be used as a single vFPGA. More general FPGA virtualization approaches for leveraging multi-FPGA systems or networked FPGA clusters shall be explored across different system stacks in future research. Additionally, although the isolation aspect of FPGA virtualization is of great importance for practical applications, this topic of research is currently under-explored. More functional, performance and fault isolation approaches for FPGA virtualization shall be explored in future research to ensure data security and system resilience.

Moreover, there has been a trend in designing specialized FPGA virtualization frameworks for commonly-used artificial intelligence and deep learning applications such as deep neural networks (DNNs) \cite{zeng2020enabling}. Most of the existing work only focus on performance and usability perspectives of the applications like enabling fast mapping from application codes to FPGA executable and creating a useful abstraction of FPGAs to compilers. For example, the solutions in \cite{zeng2020enabling,xDNN,fowers2018configurable} focus on enabling the acceleration of a large variety of DNN models using Instruction-Set-Architecture-based methods to avoid the overhead of traditional full compilation of FPGA designs. 
The existing works are either mainly focused on performance optimization \cite{xDNN,fowers2018configurable} of static DNN workload execution or they are not scalable since they do not support multi-FPGA scenarios in cloud environments \cite{zha2020virtualizing}. Moreover, the issue of isolation while sharing resources are ignored in the existing work. The work in \cite{zeng2020enabling} provides physical and performance isolation of FPGAs and tries to address functional isolation by assigning separate hardware resource pools to different users. Future research needs to address the scalability issue by supporting multi-FPGA virtualization. Furthermore, in-depth analysis and research are needed to provide better functional isolation while sharing the physical resources in multi-tenant FPGAs.

\ifCLASSOPTIONcaptionsoff
  \newpage
\fi



\bibliographystyle{IEEEtran}
\bibliography{zotero_export.bib}
%

%

\begin{IEEEbiography}[{\includegraphics[width=1in,height=1.25in,clip,keepaspectratio]{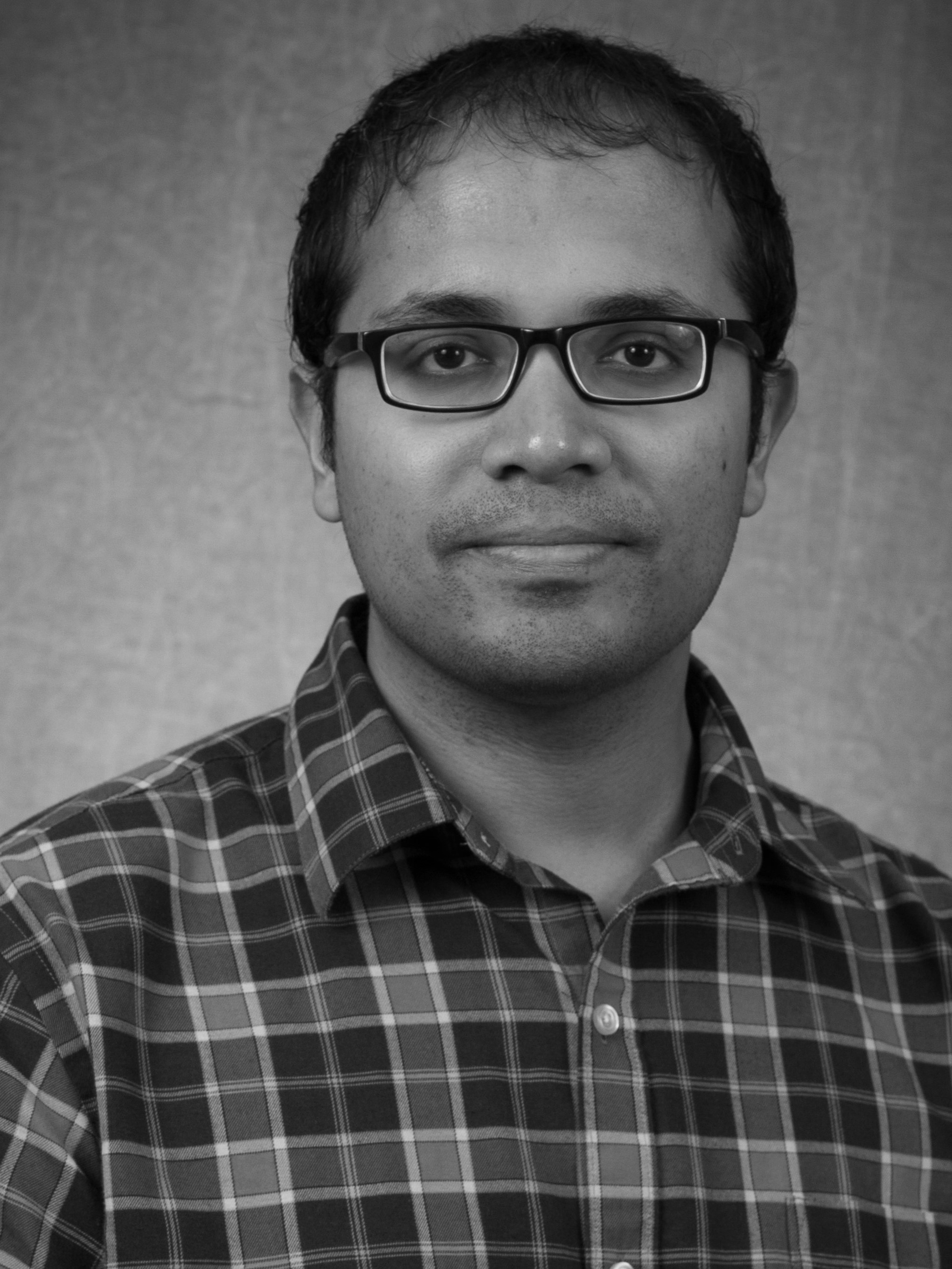}}]{Masudul Hassan Quraishi}
Masudul Hassan Quraishi received the B.Sc. Degree in Electrical Engineering from Bangladesh University of Engineering and Technology, Bangladesh in 2013 and MS degree in Computer Engineering from Arizona State University, USA in 2020. He is currently a Computer Engineering PhD student at Arizona State University, USA. His current research involves virtualization of FPGA for high performance computing. 
\end{IEEEbiography}

\begin{IEEEbiography}[{\includegraphics[width=1in,height=1.25in,clip,keepaspectratio]{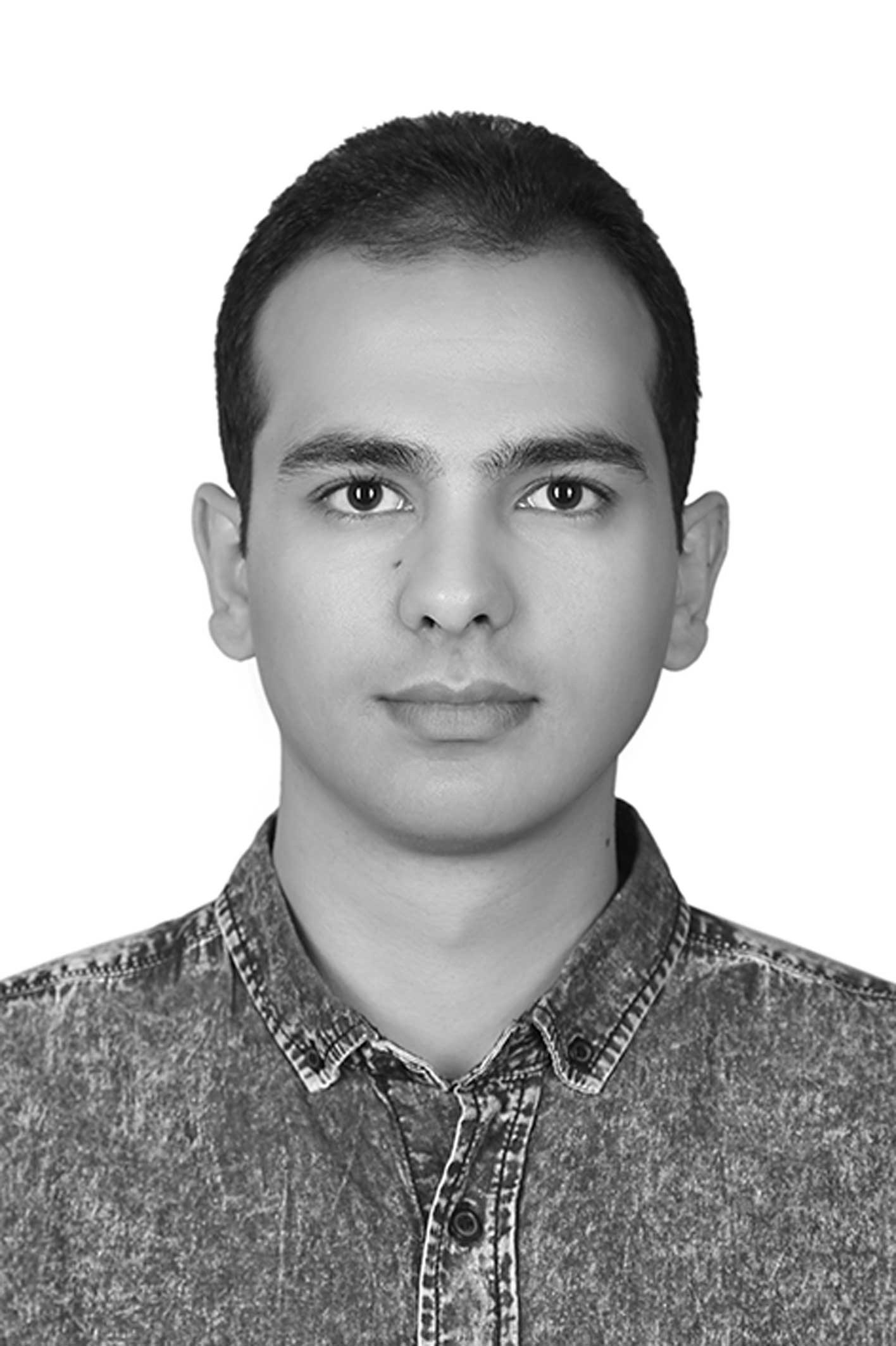}}]{Erfan Bank Tavakoli}
Erfan Bank Tavakoli received the B.Sc. degree from Iran University of Science and Technology, Tehran, Iran, in 2017, the M.Sc. degree from the University of Tehran, Tehran, Iran, in 2019, all in electrical engineering. He is currently a computer engineering Ph.D. student at Arizona State University, AZ, USA. His current research interests include hardware acceleration and deep learning.
\end{IEEEbiography}

\begin{IEEEbiography}[{\includegraphics[width=1in,height=1.25in,clip,keepaspectratio]{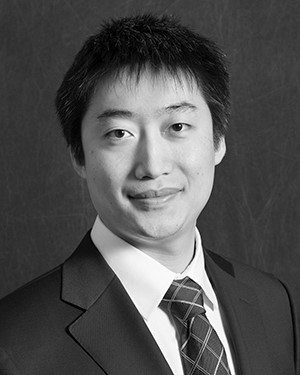}}]{Fengbo Ren}
Fengbo Ren (S’10-M’15-SM’20) received the B.Eng. Degree from Zhejiang University, Hangzhou, China, in 2008 and the M.S. and Ph.D. degrees from the University of California, Los Angeles, in 2010 and 2014, respectively, all in electrical engineering.

In 2015, he joined the faculty of the School of Computing, Informatics, and Decision Systems Engineering at Arizona State University (ASU). His Ph.D. research involved designing energy-efficient VLSI systems, accelerating compressive sensing signal reconstruction, and developing emerging memory technology. His current research interests are focused on algorithm, hardware, and system innovations for data analytics and information processing, with emphasis on bringing energy efficiency and data intelligence into a broad spectrum of today’s computing infrastructures, from data center server systems to wearable and Internet-of-things devices. He is a member of the Digital Signal Processing Technical Committee and VLSI Systems \& Applications Technical Committee of the IEEE Circuits and Systems Society.

Dr. Ren received the Broadcom Fellowship in 2012, the prestigious National Science Foundation (NSF) Faculty Early Career Development (CAREER) Award in 2017, the Google Faculty Research Award in 2018. He also received the Top 5 percent Best Teacher Awards from the Fulton Schools of Engineering at ASU in 2017, 2018, and 2019.
\end{IEEEbiography}


\vfill


\end{document}